\documentclass[prc,showpacs,superscriptaddress,twocolumn]{revtex4}

\usepackage{amsmath,bm}
\usepackage{graphicx}
\usepackage{epstopdf}
\usepackage{subfigure}
\usepackage{epsfig}
\usepackage{amsmath,amssymb,amsfonts}
\usepackage{color}
\usepackage[utf8]{inputenc}
\usepackage{verbatim}
\usepackage{array}
\usepackage{hyperref}
\usepackage{url}
\graphicspath{{figures/}} % Directory in which figures are stored

\begin{document}

\title{Energy Dependence of Elliptic Flow Ratio $v_{2}^{\text{PP}}$/$v_{2}^{\text{RP}}$ in Heavy-ion Collisions Using the AMPT Model}

\author{Shaowei Lan}
\email{shaoweilan@pdsu.edu.cn}
\affiliation{School of Electrical and Mechanical Engineering, Pingdingshan University, 467000 Pingdingshan, China}

\author{Qiuhua Liu}
%$\email{1048@pdsu.edu.cn}
\affiliation{School of Electrical and Mechanical Engineering, Pingdingshan University, 467000 Pingdingshan, China}

\author{Yong Li}
%$\email{1048@pdsu.edu.cn}
\affiliation{School of Electrical and Mechanical Engineering, Pingdingshan University, 467000 Pingdingshan, China}

\author{Shusu Shi}
\email{shiss@mail.ccnu.edu.cn}
\affiliation{Key Laboratory of Quark \& Lepton Physics (MOE) and Institute of Particle Physics, Central China Normal University, Wuhan 430079, China}

\date{\today}

%%%%%%%%%%%%%%%%%%%%%%%%%%%%%%%%%%%%%%%%%%%%%%%%%%%%%%%%%%%%%%%%%%%%%
\begin{abstract}
We present a systematic study of the elliptic flow $v_2$ relative to the participant plane (PP) and reaction plane (RP) in Au+Au collisions at $\sqrt{s_{NN}} = 7.7$--200 GeV using the AMPT model with the string melting version. The ratio $v_{2}^{\text{PP}}$/$v_{2}^{\text{RP}}$ is investigated under different hadronic cascade times (0.6 fm/$c$, 10 fm/$c$, and the maximum evolution time) and across various collision centralities. The results show that, at a fixed collision energy, the ratio exhibits negligible sensitivity to the duration of the hadronic rescattering stage, indicating that hadronic interactions have little effect on the relative difference generated by initial-state fluctuations. 
However, a strong energy dependence is observed, the ratio increases with beam energy and saturates above $\sqrt{s_{NN}} \approx 62.4$ GeV, a trend that persists across all centralities. 
These findings highlight the dominant role of the partonic phase in converting initial-state geometry fluctuations into final-state momentum anisotropy.
Conversely, at lower energies, the reduced partonic interaction strength limits this conversion efficiency, weakening the system’s ability to preserve the initial geometric information.
Our results suggest that the conversion of initial geometric fluctuations into final momentum anisotropy requires sufficient partonic interactions.
\end{abstract}

\maketitle

%%%%%%%%%%%%%%%%%%%%%%%%%%%%%%%%%%%%%%%%%%%%%%%%%%%%%%%%%%%%%%%%%%%%%
\section{Introduction}
\label{sec:introduction}
Understanding the properties of the quark-gluon plasma (QGP), a state of strongly interacting matter characterized by deconfined quarks and gluons~\cite{Braun-Munzinger:2007edi,Chen:2024aom,Bzdak:2019pkr,Shuryak:1978ij}, remains one of the goals of heavy-ion collision experiments. Relativistic heavy-ion collisions at high energies create extreme conditions where the QGP created\cite{STAR:2005gfr,BRAHMS:2004adc,PHENIX:2004vcz,PHOBOS:2004zne,STARnote0598}, allowing detailed studies of its transport properties, such as viscosity and equation of state. A crucial aspect of such investigations involves exploring how the system evolves from the initial stage through the partonic phase and eventually to the hadronic phase~\cite{Heinz:2013th,Magdy:2022ize,Bozek:2009dw,Qiu:2011iv,Niemi:2012aj,PHOBOS:2007vdf,Wang:2014boa}.

The beam energy scan (BES-I) program at the Relativistic Heavy Ion Collider (RHIC) provides a unique opportunity to study the energy dependence of QGP formation and the relative roles of partonic and hadronic interactions. In that program, Au+Au collisions were systematically measured across a wide range of center-of-mass energies ($\sqrt{s_{NN}} = 7.7 - 200$ GeV). At high energies, the system spends a longer time in the partonic phase~\cite{STAR:2015gge}, while at lower energies, the hadronic phase dominates due to the reduced initial energy density and shorter-lived partonic stage~\cite{STAR:2013ayu}. Thus, the BES-I program enables the study of the transition from partonic to hadronic dominance and its impact on final-state observables~\cite{STAR:2013cow,STAR:2017okv,STAR:2014clz,STAR:2020tga,Lan:2022rrc}.

Among these observables, the elliptic flow coefficient $v_2$ serves as a sensitive probe of the system's early-stage dynamics and its transport properties~~\cite{Romatschke:2007mq,Gale:2012rq,Song:2011qa,Shen:2020mgh}. Elliptic flow arises from the initial spatial anisotropy in the overlap region of the colliding nuclei, which is converted into final-state momentum anisotropy during the system's evolution~\cite{Sorge:1998mk,Ollitrault:1992bk,Snellings:2011sz}.
In principle, $v_2$ can be measured relative to different reference planes. The reaction plane (RP), defined by the beam direction and impact parameter vector, represents the average collision geometry, while the participant plane (PP) fluctuates event-by-event due to the random distribution of participant nucleons~~\cite{Gardim:2012yp}. Consequently, $v_2$ measured relative to the participant plane ($v_{2}^{\text{PP}}$) is typically larger than that relative to the reaction plane ($v_{2}^{\text{RP}}$), with their ratio ($v_{2}^{\text{PP}}$/$v_{2}^{\text{RP}}$) carrying information about the strength of initial-state fluctuations.
The $v_2$ presented in this paper is defined as the second fourier coefficient of particle distribution in emission azimuthal angle ($\phi$) with respect to the reaction plane angle ($\Psi^{\text{RP}}$) or participant plane angle ($\Psi^{\text{PP}}$), and can be written as~\cite{Poskanzer:1998yz,Voloshin:1994mz}:
\begin{equation}
    \frac{dN}{d\phi} \propto 1 + 2v_{2}\text{cos}(2(\phi-\Psi)).
\end{equation}

Experimental results from the STAR collaboration have reported elliptic flow for inclusive charged hadrons in Au+Au collisions across $\sqrt{s_{NN}} = 7.7 - 39$ GeV~\cite{STAR:2012och}, comparing $v_2$ obtained with respect to event planes reconstructed from different detectors, such as the TPC and BBC, which are sensitive to participant plane and reaction plane, respectively. These studies suggested a measurable difference between the two methods, underscoring the role of initial geometry fluctuations and their evolution~\cite{Renk:2014jja,Jia:2012ma,Schenke:2014tga,STAR:2021mii}.

Motivated by these observations, we perform a systematic investigation of the $v_{2}^{\text{PP}}$/$v_{2}^{\text{RP}}$ ratio using a multi-phase transport model (AMPT) with string melting mechanism, focusing on Au+Au collisions at $\sqrt{s_{NN}} = 7.7 - 200$ GeV. By varying the hadronic cascade time while fixing the partonic interaction cross section, we explore the influence of hadronic rescattering on this ratio and its dependence on collision energy and centrality~\cite{Nasim:2013fb,Zhou:2024cte}. Our study aims to explore the interplay between initial-state fluctuations and the system's evolution across the energy range covered by the RHIC BES program.

This paper is organized as follows. In Sect.~\ref{sec:model}, we briefly describe the basic features of the AMPT model. Sect.~\ref{sec:res} presents the transverse momentum ($p_T$) dependence of  $v_{2}^{\text{PP}}$, $v_{2}^{\text{RP}}$, and their ratio $v_{2}^{\text{PP}}$/$v_{2}^{\text{RP}}$, along with discussions on the centrality and collision energy dependence of the ratio. Finally, a summary of the main finding is provided in Sect.~\ref{sec:sum}.

%%%%%%%%%%%%%%%%%%%%%%%%%%%%%%%%%%%%%%%%%%%%%%%%%%%%%%%%%%%%%%%%%%%%%
\section{The AMPT model}
\label{sec:model}

We employ A Multi-Phase Transport (AMPT) model with the string melting and a parton-parton cross section 3 mb. The AMPT model has been widely used and successfully describes a variety of experimental observables, including particle production, as well as the collective flow over a broad range of collision energies~\cite{Lin:2004en,Bzdak:2014dia,Lan:2017nye,MacKay:2022uxo,Lin:2021mdn}.
The AMPT model consists of four main components: the initial conditions, partonic interactions, hadronization, and hadronic rescattering. The initial conditions are provided by the HIJING model~\cite{Wang:1991hta}, which employs Glauber nuclear geometry to simulate the spatial and momentum distributions of minijet partons from hard scattering and strings from soft processes. Partonic interactions are simulated using the Zhang's Parton Cascade (ZPC) framework~\cite{Zhang:1997ej}, which models the scatterings among partons. Once partonic interactions cease, hadronization take place though a combination of quark coalescence and string fragmentation mechanisms. The subsequent interactions among hadrons, including both elastic and inelastic scatterings, are then described by the ART (A Relativistic Transport) model~\cite{Li:1995pra}. Here, the hadronic cascade time denotes the maximum evolution time allowed for hadronic rescatterings in ART. After this time (fm/$c$), no further hadronic scatterings are evolved.

In this study the elliptic flow coefficient $v_2$ is calculated using the definition
\begin{equation}
    v_{2} = \langle \text{cos}[2(\phi - \Psi)] \rangle,
\end{equation}
where the average is taken over all particles in all events. Here, $\phi$ represents the azimuthal angle of emitted particles, and $\Psi$ denotes the azimuthal angle of the reference plane, in which can be either the reaction plane or the participant plane. In this analysis, the reaction plane angle is set to zero.
The participant plane is determined by the spatial distribution of initial partons in the transverse plane. The azimuthal angle of the second-order participant plane is defined as:
\begin{equation}
    \Psi_{2}^{\text{PP}} = \frac{\text{atan2} ( \langle r_{ini}^{2} \text{sin}(2\phi_{ini}) \rangle \langle r_{ini}^{2} \text{cos}(2\phi_{ini}) \rangle)+\pi}{2},
\end{equation}
the eccentricity with respect to second-order participant plane and reaction plane is defined as:
\begin{equation}
    \epsilon_{2}^{\text{PP}} = \frac{\sqrt{\langle r_{ini}^{2}\text{cos}(2\phi_{ini}) \rangle^{2}  + \langle r_{ini}^{2}\text{sin}(2\phi_{ini}) \rangle^{2} }}{\langle r_{ini}^{2} \rangle}
\end{equation}
\begin{equation}
   \epsilon^{\text{RP}} = \frac{\sigma_{y}^{2} - \sigma_{x}^{2}}{\sigma_{y}^{2} - \sigma_{x}^{2}}; 
   \sigma_{x}^{2} = \langle x^{2} \rangle - \langle x \rangle ^{2}, \sigma_{y}^{2} = \langle y^{2} \rangle - \langle y \rangle ^{2};
\end{equation}
where $r_{ini}$ and $\phi_{ini}$ are the polar coordinate position of each nucleon and the average brackets is density weighted in the initial state,  the $x$ and $y$ are the position of participant nucleons.

%%%%%%%%%%%%%%%%%%%%%%%%%%%%%%%%%%%%%%%%%%%%%%%%%%%%%%%%%%%%%%%%%%%%%
\section{Results and discussions}
\label{sec:res}

In the following sections, we present a comprehensive analysis of the elliptic flow coefficient $v_2$ of charged hadrons relative to both the participant plane (PP) and the reaction plane (RP) using the AMPT model with the string melting version. Specifically, we show the transverse momentum ($p_T$) dependence of $v_{2}^{\text{PP}}$ and $v_{2}^{\text{RP}}$ for charged hadrons across different collision centralities and collision energies. The ratio $v_{2}^{\text{PP}}$/$v_{2}^{\text{RP}}$ is also analyzed as a function of $p_T$, and its centrality dependence is quantified using the linear fits to the $p_T$ differential results. 
Furthermore, we investigate the effect of hadronic rescattering on the ratio at each collision energy. Finally, we provide a detailed discussion of the energy and centrality dependence of the $v_2$ ratio.

\begin{figure*}[!htb]
\centering
\centerline{\includegraphics[width=1.0\textwidth]{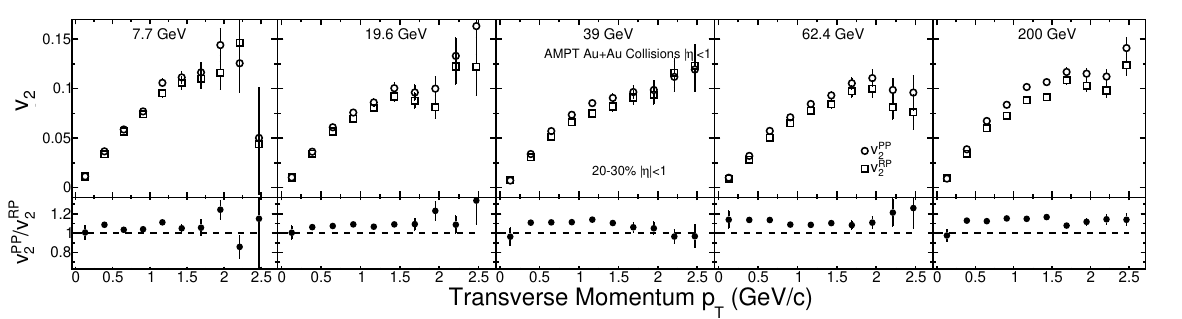}}
\caption{The top panels show $v_2$ of charged hadrons as a function of $p_T$ for Au+Au 20-30\% collisions at $\sqrt{s_{NN}} = 7.7,19.6, 39, 62.4,$ and 200 GeV from AMPT model at midrapidity ($|\eta|<$1). The results are shown with a parton-parton cross section of 3 mb and a hadronic cascade time evolved up to the maximum evolution time. The open circles and open squares represent the results of $v_{2}^{\text{PP}}$ and $v_{2}^{\text{RP}}$, respectively. The bottom panels show the ratio of $v_{2}^{\text{PP}}$/$v_{2}^{\text{RP}}$ as a function of $p_T$ for all centralities.}
\label{fig1}
\end{figure*}

Following the experimental analysis procedures~\cite{STAR:2013ayu}, the collision centrality is determined from the reference multiplicity defined as the number of charged pions, kaons, and protons within the pseudorapidity range $|\eta|<$0.5 and transverse momentum range $0.2 < p_{T} < 2.0 $ GeV/$c$. Centrality classes~\cite{Kharzeev:2000ph} are determined by percentile cuts on the reference multiplicity distribution from the minimum-bias AMPT events for each energy.
The top panels of Figure~\ref{fig1} display the transverse momentum ($p_T$) dependence of the $v_2$ for charged hadrons in 20-30\% central Au+Au collisions at mid-rapidity ($|\eta|<$1), for collision energies of 7.7, 19.6, 39, 62.4, and 200 GeV, shown from left to right. The open circles and open squares represent the results of $v_2$ relative to the participant plane ($v_{2}^{\text{PP}}$) and the reaction plane ($v_{2}^{\text{RP}}$), respectively. Both $v_{2}^{\text{PP}}$ and $v_{2}^{\text{RP}}$ exhibit a similar $p_T$ dependence, increasing with $p_T$ and reaching a maximum at higher $p_T$ region. Notably, $v_{2}^{\text{PP}}$ is consistently larger than $v_{2}^{\text{RP}}$ across the entire $p_T$ range, in agreement with theoretical expectations. Moreover, the difference between $v_{2}^{\text{PP}}$ and $v_{2}^{\text{RP}}$ appears to exhibit an energy dependence.

To quantify the difference between $v_{2}^{\text{PP}}$ and $v_{2}^{\text{RP}}$ across collision energies, we evaluate their ratio $v_{2}^{\text{PP}}$/$v_{2}^{\text{RP}}$ in each $p_{T}$ bin, as shown by the solid circles in the lower panels of Figure~\ref{fig1}.
It is important to note that both $v_{2}^{\text{PP}}$ and $v_{2}^{\text{RP}}$ are calculated from the same set of particles within each event, which inevitably introduces statistical correlations between the two observables. A simple propagation of uncertainties or a straightforward fitting procedure would therefore overestimate the true errors of the ratio. To properly address this issue, we employ a bootstrap resampling technique to estimate the statistical uncertainties of $v_{2}^{\text{PP}}$/$v_{2}^{\text{RP}}$ throughout this work~\cite{Efron:1986hys}.
The bootstrap method provides a robust, non-parametric way of evaluating uncertainties by repeatedly resampling the particle ensemble and recalculating the observables. This approach naturally preserves the inherent correlations between $v_{2}^{\text{PP}}$ and $v_{2}^{\text{RP}}$, while avoiding biases that arise in conventional error propagation.
A clear energy dependence is observed that the ratio $v_{2}^{\text{PP}}$/$v_{2}^{\text{RP}}$ increases with increasing collision energy. And this behavior is also consistent with the experimental measurements from the STAR BES-I experiment~\cite{STAR:2012och}.

\begin{figure*}[!htb]
\centering
\centerline{\includegraphics[width=1.0\textwidth]{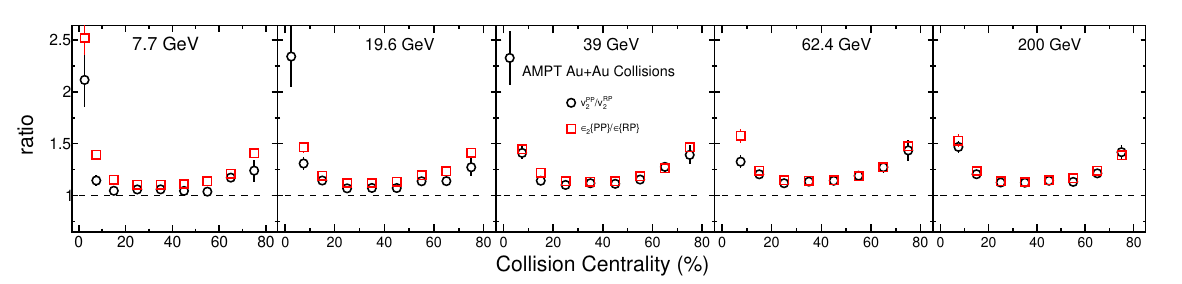}}
\caption{The ratio of elliptic flow coefficients $v_{2}^{\text{PP}}$/$v_{2}^{\text{RP}}$ and corresponding eccentricity ratio $\epsilon_{2}{\text{(PP)}} / \epsilon{\text{(RP)}}$ of charged hadrons as a function of collision centrality for Au+Au collisions at $\sqrt{s_{NN}} = 7.7,19.6, 39, 62.4,$ and 200 GeV from the AMPT model at midrapidity ($|\eta|<$1). The results are shown for a parton-parton cross section of 3 mb and a hadronic cascade time evolved up to the maximum evolution time. The black open circles represent the $v_{2}^{\text{PP}}$/$v_{2}^{\text{RP}}$ ratios, while the red open squares denote the $\epsilon_{2}{\text{(PP)}} / \epsilon{\text{(RP)}}$ ratios.}
\label{fig2}
\end{figure*}

The centrality dependence of the ratio $v_{2}^{\text{PP}}$/$v_{2}^{\text{RP}}$ is shown by black open circles in Figure~\ref{fig2}. At a given collision energy, the ratio exhibits a non-monotonic centrality dependence that it reaches a minimum in mid-central collisions and increases toward both central and peripheral collisions. To further investigate the role of initial-stage fluctuations, we also calculate the centrality dependence of the ratio  $\epsilon_{2}^{\text{PP}} / \epsilon^{\text{RP}}$, shown as the red open squares. 
Both ratios are consistently larger than unity and display similar centrality and energy dependence. 
These observations indicate that the effects of initial fluctuations on elliptic flow are influenced by both the initial eccentricity (or system size) and/or the dynamical evolution mechanisms that vary with beam energy.

To further investigate the impact of hadronic interactions on the initial-state fluctuations in elliptic flow measurements, we performed AMPT simulations of Au+Au collisions with a fixed parton-parton cross section of 3 mb. The hadronic cascade time was varied among 0.6 fm/$c$, 10 fm/$c$, and 30 fm/$c$, where 30 fm/$c$ corresponds to the maximum evolution time allowed for hadronic rescatterings in AMPT at these energies. A larger hadronic cascade time corresponds to stronger hadronic rescattering effects. In the following, this setting is referred to simply as the maximum evolution time.

\begin{figure*}[!htb]
\centering
\centerline{\includegraphics[width=0.8\textwidth]{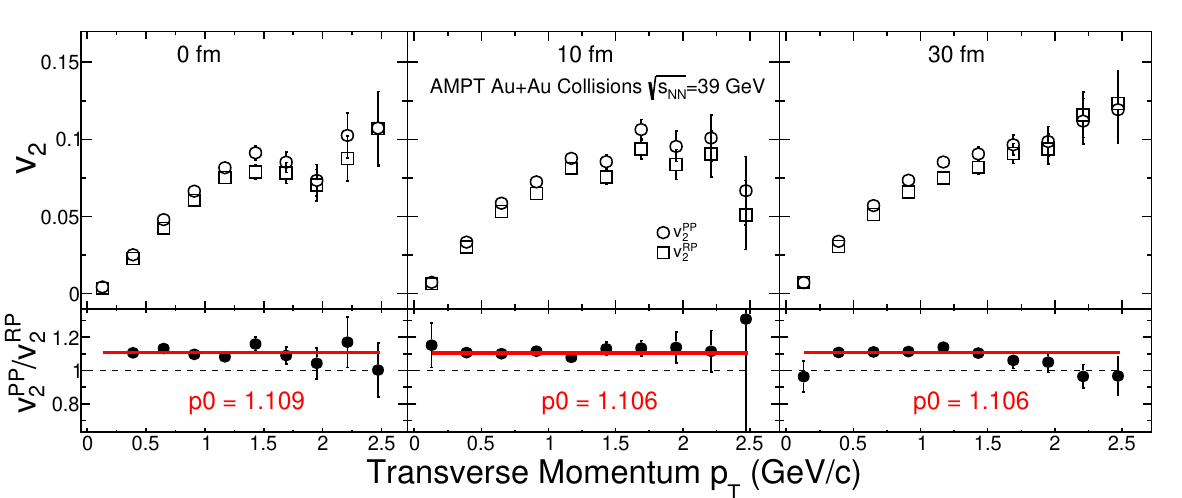}}
\caption{The top panels show $v_2$ of charged hadrons as a function of $p_T$ for Au+Au 20-30\% collisions at $\sqrt{s_{NN}} = 39$ GeV from AMPT model at midrapidity ($|\eta|<$1). The results are shown for a parton-parton cross section of 3 mb and three different values of hadronic cascade time periods. The open circles and open squares represent the results of $v_{2}^{\text{PP}}$ and $v_{2}^{\text{RP}}$, respectively. The bottom panels show the ratio of $v_{2}^{\text{PP}}$/$v_{2}^{\text{RP}}$ as a function of $p_T$ for the three different values of hadronic cascade time. The red line represents the constant fit to the ratios.}
\label{fig3}
\end{figure*}

Figure~\ref{fig3} shows the $p_{T}$ dependence of $v_{2}^{\text{PP}}$ and $v_{2}^{\text{RP}}$ for charged hadrons in 20-30\% central Au+Au collisions at $\sqrt{s_{NN}} = 39$ GeV. The results are presented for hadronic cascade times of 0.6 fm/$c$, 10 fm/$c$, and the maximum evolution time, corresponding to the left, middle, and right panels, respectively. The lower panels display the ratio $v_{2}^{\text{PP}}$/$v_{2}^{\text{RP}}$ as a function of $p_{T}$, shown by the black solid circles. At each cascade time, $v_{2}^{\text{PP}}$ and $v_{2}^{\text{RP}}$ exhibit similar $p_{T}$ dependence, and the ratio remains above unity across the entire $p_{T}$ range, with $v_{2}^{\text{PP}}$ consistently larger than $v_{2}^{\text{RP}}$. 
A constant fit to the ratio (red line) shows that the fitted values remain nearly unchanged with different hadronic cascade times.

It is also observed that within the given centrality range, the ratio $v_{2}^{\text{PP}}$/$v_{2}^{\text{RP}}$ remains nearly constant as the hadronic cascade time increases. Although the absolute magnitudes of $v_{2}^{\text{PP}}$ and $v_{2}^{\text{RP}}$ increase with hadronic evolution, their changes are comparable, leaving the ratio essentially unaffected. 
This stability indicates that while the hadronic stage modifies the overall magnitude of $v_2$, it has little impact on the relative difference between $v_{2}^{\text{PP}}$ and $v_{2}^{\text{RP}}$. The ratio is therefore likely established during the partonic phase.

\begin{figure*}[!htb]
\centering
\centerline{\includegraphics[width=0.6\textwidth]{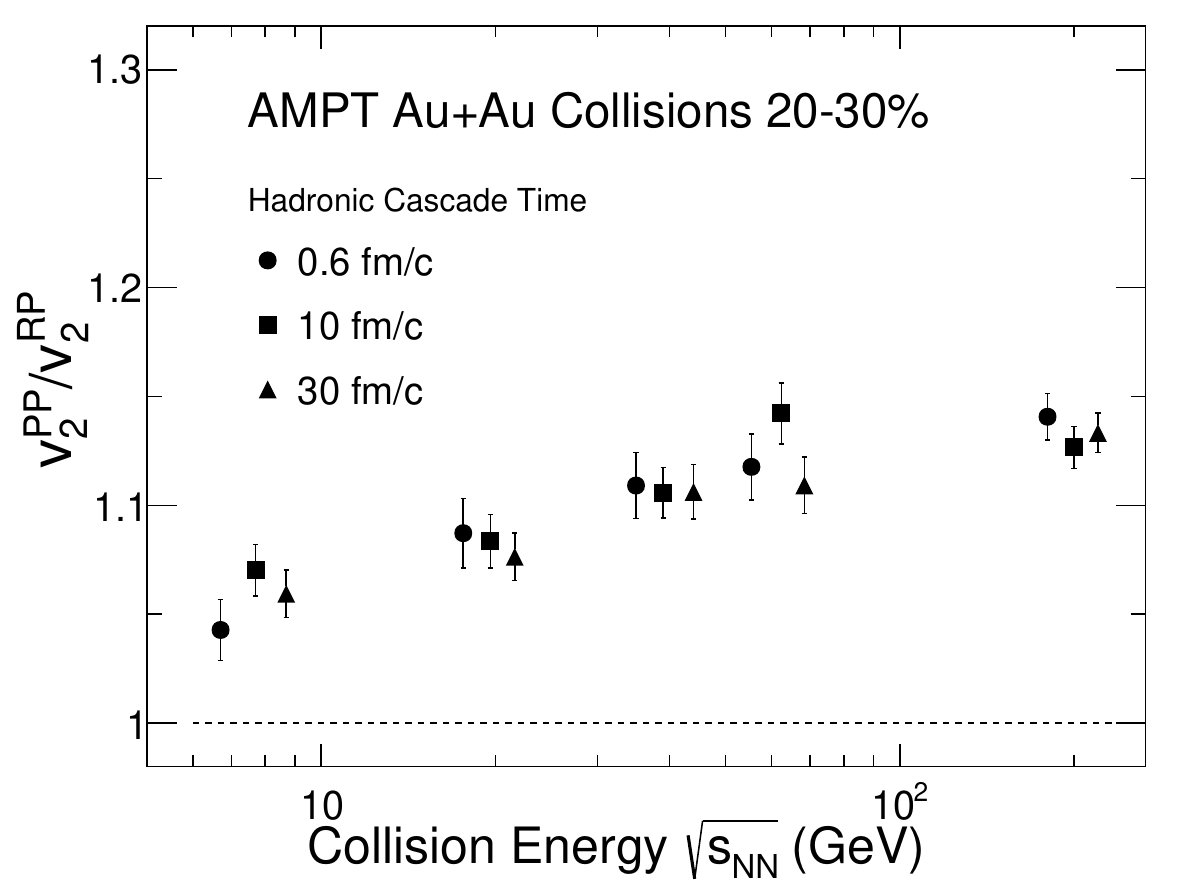}}
\caption{The ratio of $v_{2}^{\text{PP}}$/$v_{2}^{\text{RP}}$ of charged hadrons as a function of collision energy for Au+Au 20-30\% collisions from AMPT model. The results are shown for a parton-parton cross section of 3 mb and three different value of hadronic cascade time of 0.6 fm/$c$, 10 fm/$c$, and 30 fm/$c$, respectively.}
\label{fig4}
\end{figure*}

Figure~\ref{fig4} presents the collision energy dependence of the ratio $v_{2}^{\text{PP}}$/$v_{2}^{\text{RP}}$ for charged hadron in 20-30\% central collisions, along with the effects of hadronic rescattering at each collision energy. The results are obtained with a parton-parton cross section of 3 mb and three different values of hadronic cascade time of 0.6 fm/$c$, 10 fm/$c$, and the maximum evolution time, shown by the solid circles, solid squares, and solid triangles, respectively.
As shown in the figure, at a given collision energy, the ratio $v_{2}^{\text{PP}}$/$v_{2}^{\text{RP}}$ remains nearly constant when varying the hadronic cascade time, indicating that hadronic rescatterings have a negligible effect on the ratio. In contrast, a pronounced collision energy dependence is observed that the ratio increases with collision energy and tends to saturate between 62.4 and 200 GeV, suggesting a possible upper limit of the partonic interaction effect on the initial-state fluctuations of elliptic flow. These results imply that the contribution of fluctuations from the partonic phase becomes saturated at high energies.

The negligible sensitivity of the $v_2$ ratio to the hadronic cascade time suggests that the difference between $v_{2}^{\text{PP}}$ and $v_{2}^{\text{RP}}$ is primarily established during the early partonic stage of the collision. The preservation of initial geometry information requires a sufficiently long-lived and dominant partonic stage. As the collision energy decreases, both the duration and significance of the partonic phase diminish, while hadronic interactions become increasingly important. As a result, the medium's capacity to preserve the initial-stage geometry is weakened, leading to changes in the $v_2$ values relative to different reference planes.

\begin{figure*}[!htb]
\centering
\centerline{\includegraphics[width=0.6\textwidth]{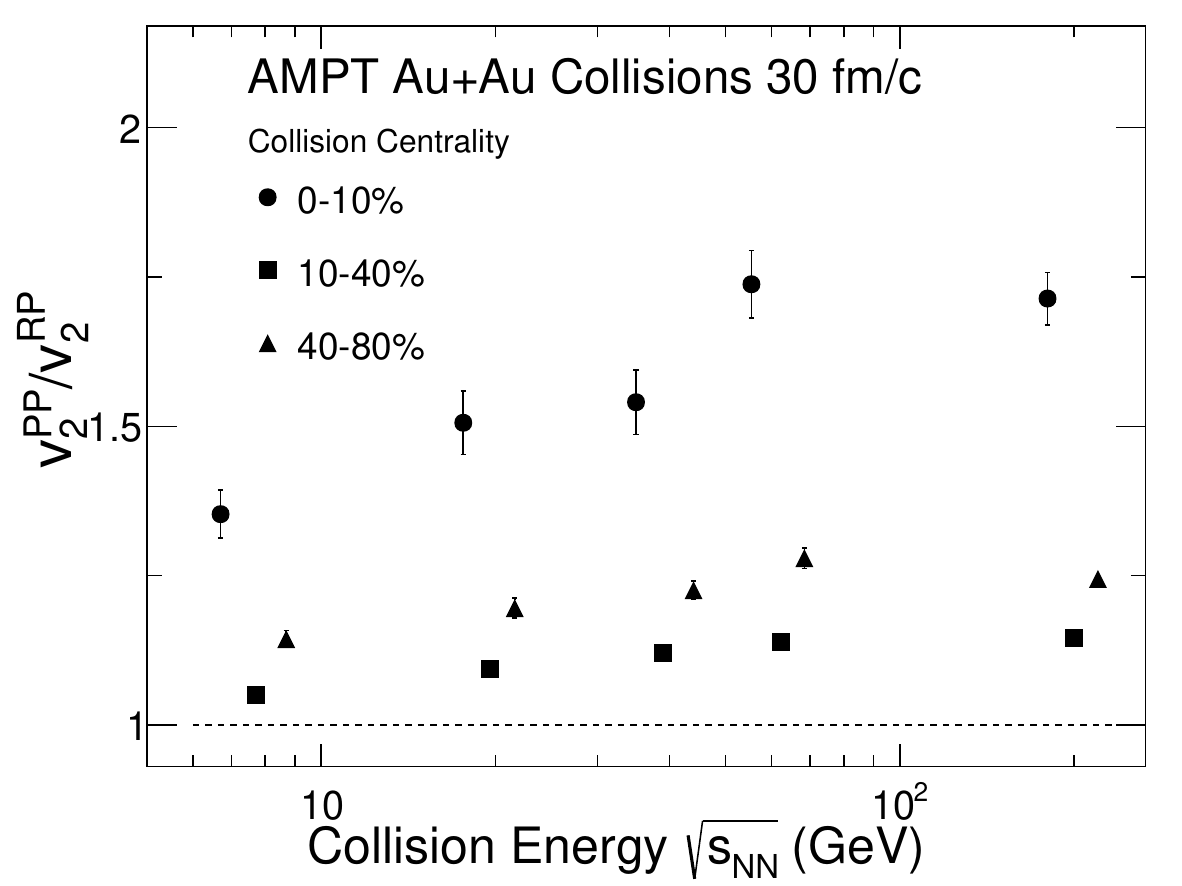}}
\caption{The ratio of $v_{2}^{\text{PP}}$/$v_{2}^{\text{RP}}$ of charged hadrons as a function of collision energy for Au+Au collisions at 0-10\%, 10-40\% and 40-80\% from AMPT model. The results are shown for a parton-parton cross section of 3 mb and a hadronic cascade time of 30 fm/$c$.}
\label{fig5}
\end{figure*}

Similarly, Figure~\ref{fig5} shows the collision energy dependence of the ratio $v_{2}^{\text{PP}}$/$v_{2}^{\text{RP}}$ for charged hadrons, along with its centrality dependence at each energy. The results are performed with a parton-parton cross section of 3 mb and a hadronic cascade evolved to the maximum evolution time. The solid circle, solid triangles, and solid squares correspond to the 0-10\%, 10-40\%, and 40-80\% centrality classes, respectively.
A significant collision energy dependence is observed across all centrality classes that the ratio increases with increasing collision energy and tends to saturate above $\sqrt{s_{NN}} = 62.4$ GeV. In addition, the ratio exhibits a clear centrality dependence at each collision energy. Specifically, it reaches the largest value in the most central collisions, decreases gradually towards peripheral collisions, and shows a distinct minimum in mid-central collisions.

This pronounced centrality dependence of the ratio $v_{2}^{\text{PP}}$/$v_{2}^{\text{RP}}$ arises from the interplay between the initial geometric eccentricity and fluctuations of the participant plane. In most central collisions, the overlap region of the colliding nuclei is nearly symmetric, yielding in a small geometric eccentricity. In this regime, event-by-event fluctuations dominate, leading to a significant deviation between the participant and reaction planes. As a result, $v_{2}^{\text{PP}}$ becomes significantly larger than $v_{2}^{\text{RP}}$, producing a high ratio.
In mid-central collisions, where the initial geometric eccentricity is well-defined, the relative importance of fluctuations is reduced. This improves the alignment between the participant and reaction planes, leading to a lower ratio approaching unity.
In peripheral collisions, although the geometric eccentricity remains finite, the reduced number of participant nucleons enhances statistical fluctuations, which again increases the ratio.
Overall, the observed centrality dependence of the ratio $v_{2}^{\text{PP}}$/$v_{2}^{\text{RP}}$ reflects the competition between the geometric shape of the initial overlap region and the magnitude of initial-state fluctuations, both of which vary systematically with collision centrality.

%%%%%%%%%%%%%%%%%%%%%%%%%%%%%%%%%%%%%%%%%%%%%%%%%%%%%%%%%%%%%%%%%%%%%
\section{Summary}
\label{sec:sum}

In summary, we have systematically investigated the ratio of elliptic flow coefficients $v_{2}^{\text{PP}}$/$v_{2}^{\text{RP}}$ for charged hadrons in Au+Au collisions over a broad range of collision energies using the AMPT model with string mealting version. Simulations are performed at $\sqrt{s_{NN}} = 7.7, 19.6, 39, 62.4$ and 200 GeV, with a fixed parton-parton cross section of 3 mb and hadronic cascade times of 0.6 fm/$c$, 10 fm/$c$, and the maximum evolution time.

The results show that the $v_2$ ratio is essentially insensitive to the hadronic cascade time aross all collision energies, indicating that it is primary determined during the early partonic phase of the system evolution.
A pronounced energy dependence is observed that the ratio increases with collision energy and saturate above $\sqrt{s_{NN}} \approx$ 62.4 GeV. This behavior suggests that the conversion of initial geometric fluctuations into final momentum anisotropy requires sufficient partonic interactions.

A strong centrality dependence is also observed. The ratio reaches its maximum in the most central collisions due to the dominance of initial-state fluctuations in nearly symmetric overlap regions. In mid-central collisions, the geometric eccentricity is well-defined and the ratio reaches its minimum. In peripheral collisions, the reduced number of participating nucleon increases the relative contribution of statistical fluctuations, causing the ratio to rise again.

These findings demonstrate that the ratio $v_{2}^{\text{PP}}$/$v_{2}^{\text{RP}}$ provides a sensitive probe of the interplay between initial geometry and the subsequent dynamical evolution of the medium, thereby offering valuable insights into the relative role of partonic and hadronic phases across different collision energies and centralities.

\vspace*{-0.2in}
\acknowledgments
This work was supported by the Doctoral Scientific Research Foundation of Pingdingshan University (PXY-BSQD-2023016), and the Natural Science Foundation of Henan under contract No. 252300420921, the National Key Research and Development Program of China under Contract No. 2022YFA1604900, and the National Natural Science Foundation of China (NSFC) under contract No. 12175084.

%%%%%%%%%%%%%%%%%%%%%%%inroduction%%%%%%%%%%%%%%%%
%\bibliographystyle{unsrt}
\bibliography{ref.bib}

%%%%%%%%%%%%%%%%%%%%%%bibliographyend%%%%%%%%%%%%%

\end{document}